\documentclass[twocolumn,english]{article}
\usepackage{authblk}
\usepackage{etoolbox}
\usepackage{lmodern}
\usepackage{subcaption}
\usepackage{graphicx}
\usepackage{url}
\PassOptionsToPackage{hyphens}{url}
\usepackage{multirow}
\usepackage{wasysym}
\usepackage{marvosym}
\usepackage{csquotes}

\makeatletter 
\def\@normalsize{\@setsize\normalsize{10pt}\xpt\@xpt
\abovedisplayskip 10pt plus2pt minus5pt\belowdisplayskip \abovedisplayskip
\abovedisplayshortskip \z@ plus3pt\belowdisplayshortskip 
6pt plus3pt minus3pt\let\@listi\@listI}
\def\subsize{\@setsize\subsize{12pt}\xipt\@xipt}
\def\section{\@startsection {section}{1}{\z@}
	{1.8ex plus 1ex minus .2ex} 
	{1.2ex plus .2ex \@afterindentfalse}
	{\large\bf}}
\def\subsection{\@startsection {subsection}{2}{\z@}
	{1.3ex plus 1ex}
	{.8ex plus .2ex \@afterindentfalse}
	{\subsize\bf}}
\def\paragraph{\@startsection {paragraph}{4}{\z@}
	{1.8ex plus .3ex}
	{-1em  \@afterindentfalse}
	{\normalsize\bf}}

\let\OLDthebibliography\thebibliography
\renewcommand\thebibliography[1]{
	\OLDthebibliography{#1}
	\setlength{\parskip}{0pt}
	\setlength{\itemsep}{0pt plus 0.3ex}
}

\expandafter\def\expandafter\UrlBreaks\expandafter{\UrlBreaks% save the current one
	\do\a\do\b\do\c\do\d\do\e\do\f\do\g\do\h\do\i\do\j%
	\do\k\do\l\do\m\do\n\do\o\do\p\do\q\do\r\do\s\do\t%
	\do\u\do\v\do\w\do\x\do\y\do\z\do\A\do\B\do\C\do\D%
	\do\E\do\F\do\G\do\H\do\I\do\J\do\K\do\L\do\M\do\N%
	\do\O\do\P\do\Q\do\R\do\S\do\T\do\U\do\V\do\W\do\X%
	\do\Y\do\Z\do\*\do\-\do\~\do\'\do\"\do\-}%

\patchcmd{\@maketitle}{\LARGE \@title}{\fontsize{16}{19.2}\selectfont\@title}{}{}
\makeatother

\title{Using Vision Videos in a Virtual Focus Group:\\ Experiences and Recommendations}

\author[1]{Oliver Karras}
\author[2]{Svenja Polst}
\author[2]{Kathleen Späth}

\affil[1]{Leibniz Universität Hannover, Software Engineering Group, 30167 Hannover, Germany}
\affil[ ]{Email: oliver.karras@inf.uni-hannover.de}
\affil[2]{Fraunhofer IESE, 67663 Kaiserslautern, Germany}
\affil[ ]{Email: \{svenja.polst, kathleen.spaeth\}@iese.fraunhofer.de}
\date{}

\begin{document}
\maketitle
\section{Introduction}
Facilitated meetings are an established practice for the requirements engineering activities \textit{elicitation} and \textit{validation} \cite{Wagner.2019}. Focus groups are one well-known technique to implement this practice. Several researchers \cite{Bennaceur.2016, Darby.2018, Rodden.2013} already reported the successful use of vision videos to stimulate active discussions among the participants of on-site focus groups, e.g., for validating scenarios and eliciting feedback. These vision videos show scenarios of a system vision. In this way, the videos serve all parties involved as a visual reference point to actively disclose, discuss, and align their mental models of the future system to achieve shared understanding \cite{Karras.2020a}. In the joint project TrUSD\footnote{\scriptsize{\url{https://www.trusd-projekt.de/}}}, we had planned to conduct such an on-site focus group using a vision video to validate a scenario of a future software tool, the so-called \textit{Privacy Dashboard}\footnote{\scriptsize{\url{https://www.trusd-projekt.de/wp/motivation-loesungsidee/}}}. However, the COVID-19 pandemic and its associated measures led to an increase in home and remote working, which also affected us. Therefore, we had to replan and conduct the focus group virtually. In this paper, we report about our experiences and recommendations for the use of vision videos in virtual focus groups.

\section{Setting of the Virtual Focus Group}
The goal of the virtual focus group was to validate whether the current \textit{Privacy Dashboard} concepts reflect the users' needs and are consistent with their workflows. Therefore, we focused on use cases and not on design decisions. In a previous on-site focus group, we presented the \textit{Privacy Dashboard} as a slide show consisting of static images of mockups. We had the impression that the participants of the on-site focus group had difficulties following the moderator's verbal descriptions of the scenario. For this reason, we decided to present the scenario as a vision video depicting animations such as typing and cursor movements. We produced the vision video by using the \textit{Mockup Recorder} \cite{Karras.2017a}. The \textit{Mockup Recorder} allows producing vision videos of scenarios by interacting with static images of mockups without any implementation \cite{Karras.2017a}.

The virtual focus group took place on 05$^{th}$ August 2020 with two future users, one moderator, and two researchers as observers. The procedure of the virtual focus group was as follows: (1) The focus group started with an introduction to the idea behind the \textit{Privacy Dashboard}. For this purpose, a video was shown that originally had been produced for another workshop, hereinafter referred to as the introduction video. Then, (2) the moderator played the vision video in the \textit{Mockup Recorder} that was presented to the participants via screen sharing. The vision video showed navigation through the \textit{Privacy Dashboard} and the specified scenario on several mockups. The moderator commented on the vision video live. Afterward, (3) the participants watched the vision video a second time on their own without the explanations of the moderator. Next, (4) the moderator played parts of the vision video again to discuss them in the group. One of the observers collected feedback on the mockups and their interaction processes. The participants also provided feedback to the virtual focus group and vision video. In the end, (5) the participants completed a questionnaire regarding the application and quality of the vision video.

\section{Experiences and Recommendations}
In the following, we report about our experiences and recommendations for the use of a vision video in a virtual focus group to validate a scenario.

\paragraph{Video Production.} 
Despite the prototypical implementation of the \textit{Mockup Recorder} and its moderate usability, we were able to produce the 2-minute vision video in about 90 minutes. The current version of the \textit{Mockup Recorder} produces only vision videos without sound. However, this function is under development. In the third step, the participants asked several questions about the vision video while watching it alone. Although they watched the vision video with the moderator and her explanations right before, they lacked explanations on the second viewing. We recommend to include texts in the vision video or an audio track so that participants have more explanations when they watch the vision video alone.

In the fifth step, we investigated the quality of the vision video with the questionnaire. The results show that the participants found the overall quality of the vision video to be good. Despite the currently missing function to add sound, we recommend the \textit{Mockup Recorder} to quickly produce a vision video at moderate costs and with sufficient quality.

\paragraph{Video Streaming.} 
As the communication platform, we used \textit{Microsoft Teams}\footnote{\scriptsize{\url{https://www.microsoft.com/de-de/microsoft-365/microsoft-teams/group-chat-software}}}. In the first step, the moderator used the introduction video to explain the idea behind the \textit{Privacy Dashboard}. During playback, however, the platform did not transmit the sound of the introduction video, and the moderator was unable to solve the problem immediately. The video transmission was also subject to some delays that caused the image to pause. We counteracted these issues by offering both videos (introduction video and vision video) as downloadable files for the participants so that they could watch the videos locally on their computers. Based on this experience, we recommend to offer the videos used as separate files for download.

\paragraph{Video Presentation.}
In the second step, it was a challenge for the moderator to adapt her live explanations to the pace of the video. However, instead of adding sound separately, we consciously decided that the moderator comment on the vision video live to save effort. In the vision video, each mockup is shown for as long as the associated interaction process lasts. We assumed that the realistic pace would be necessary for a realistic representation of the \textit{Privacy Dashboard}. However, the duration that some mockups were shown was sometimes too short, i.e., less than 5 seconds, to perceive and explain the details. Furthermore, we did not get the impression that a realistic representation was important for the participants. For example, typing texts in the vision video was not displayed as smooth as in reality since the \textit{Mockup Recorder} simulates the interaction processes to generate the vision video. However, no participant was bothered by it. Therefore, we recommend to extend the duration of the interaction process to stay longer on the individual mockups. We also expected that the participants would use a large enough screen to participate in the virtual focus group. Although the participants had larger screens at their disposal, they only used small laptop screens. We assume that the participants deliberately chose the smaller screen to look directly into the integrated webcam. However, the small screens made it difficult for the participants to recognize the details in the vision video. We recommend to actively ask the participants to use a sufficiently large screen and the full-screen view.

In the fourth step, we planned to show the vision video section by section and discuss each section with the participants. However, the comments and questions of the participants did not follow the sequence of the vision video. The moderator had to jump back and forth several times in the video. The moderator used the \textit{Mockup Recorder} for the presentation of the vision video. The \textit{Mockup Recorder} allowed her to select every single static mockup in the video and start the video from the selected mockup. This functionality was useful since it would have been much more difficult to find the corresponding mockup in the mere video. Another solution could be to divide the scenario into smaller units of meaning and produce one vision video for each unit. In this way, the moderator has more control over the focus group making it easier to elicit specific feedback from the participants on the particular video. We recommend to either use a software tool with similar functionality as the \textit{Mockup Recorder}, prepare several short videos according to the units of meaning in the scenario, or prepare an additional slide show as in our on-site focus group.

\section{Future Work}
We presented our experiences and recommendations for the use of vision videos in virtual focus groups. So far, we only conducted a single focus group with two participants. We plan to repeat the validation of scenarios using vision videos in virtual focus groups with more participants. In this way, we hope to gain further insights into the technique of using vision videos in virtual focus groups. Based on our observations, we also want to investigate the hypothesis that the use of a vision video leads to more feedback on functionality and less feedback on design decisions.

\section*{Acknowledgment}\label{sec:acknowledgment}
This work was supported by the German Ministry of Education and Research (BMBF) under Reference No.: $16$KIS$0896$K, $16$KIS$0898$ and the Deutsche Forschungsgemeinschaft (DFG) under Grant No.: $289386339$, ($2017$ -- $2019$).

\bibliographystyle{abbrvdin} 
{\scriptsize \bibliography{references}}
%footnotesize
\end{document}